\documentclass[12pt,a4paper]{article}
\usepackage{geometry}
\usepackage{graphicx}
\usepackage{color}
\usepackage[colorlinks=true,citecolor=black,linkcolor=black,urlcolor=black]{hyperref}
\usepackage[square,numbers,comma,colon,sort&compress]{natbib}
\usepackage{subcaption}
\usepackage{amssymb}
\usepackage{amscd}
\usepackage{amsmath}
\usepackage{tensor}
\usepackage{tikz}
\usetikzlibrary{patterns,intersections,calc,decorations.markings}

\definecolor{MS}{rgb}{1,0,0}

\begin{document}
\begin{titlepage}

{\hbox to\hsize{\hfill 
April 2021
}}

\bigskip

\bigskip 

\bigskip

\bigskip 

\vspace{3\baselineskip}

\begin{center}
{\bf \Large 
Probing Quadratic Gravity with Binary Inspirals}

\bigskip

\bigskip

\bigskip

{\bf Yunho Kim, Archil Kobakhidze and Zachary S. C. Picker \\ }

\smallskip

{ \small \it 
School of Physics, The University of Sydney, NSW 2006, Australia \\
}

\bigskip
 
\bigskip

\bigskip

{\large \bf Abstract}

\end{center}
\noindent
In this paper, we study gravitational waves generated by binary systems within an extension of General Relativity which is described by the addition of quadratic in curvature tensor terms to the Einstein-Hilbert action. Treating quadratic gravity as an effective theory valid in the low energy/curvature regime, we argue that reliable calculations can be performed in the early inspiral phase, and furthermore, no flux of additional massive waves can be detected. We then compute the Yukawa suppressed dipole-like corrections to the post-Newtonian (PN) expansion of the standard waveform. By confronting these theoretical calculations with available experimental data, we constrain both unknown parameters of quadratic gravity to be $0 \leq \gamma \, \lesssim 5.9\cdot 10^{76}$, and $-\frac{\gamma}{4} \leq \beta \, \lesssim 9.8\cdot 10^{75} - \frac{\gamma}{4}$.

\bigskip
 
\bigskip

\bigskip

 \end{titlepage}

\section{Introduction}\label{sec1}
The observations of gravitational waves from compact binary systems provide a testing ground for various speculative hypotheses regarding extensions of General Relativity (GR) \cite{LIGOScientific:2019fpa} (see also \cite{Berti:2018cxi,Tahura:2018zuq}, and the references therein). For example, it has been demonstrated in \cite{Kobakhidze:2016cqh} that remarkably strong constraints on noncommutative spacetime follow from the observed signals. An extension of GR obtained by the addition of higher curvature terms to the simplest diffeomorphism-invariant Einstein-Hilbert action is arguably the least speculative modification. In fact,   
the quantum nature of matter makes such a modification of the Einstein theory of gravitation 
unavoidable. In particular, one-loop renormalisation in the semi-classical gravity regime leads to the gravitational action with additional quadratic in curvature tensor terms \cite{Utiyama:1962sn}:

\begin{equation}\label{oriact1}
    S = \int d^4x \sqrt{-g} \left[\frac{R}{2\kappa} + \beta R^2 + \gamma R^{\mu \nu} R_{\mu \nu} \right],
\end{equation}
where $\kappa = 8 \pi G=1/M_P^2$ [$M_P\approx 2.4\cdot 10^{18}$ GeV is the reduced Planck mass in natural units], and $\beta$, and $\gamma$ are the dimensionless parameters\footnote{A possible term proportional to the square of the Riemann tensor, $R_{\mu\nu\rho\sigma}R^{\mu\nu\rho\sigma}$, in (\ref{oriact1}) can be eliminated using the Gauss-Bonnet identity. In a spacetime with boundaries there is an additional term proportional to the covariant d'Alambertian of the Ricci scalar \cite{deBerredoPeixoto:2004if}, $\Box R$, which is not relevant for our discussion and we omit it in (\ref{oriact1}).}. Due to the nonrenormalisability of gravity, these parameters cannot be computed theoretically and must be inferred from experiments. Furthermore, higher loops generate an infinite series of higher curvature terms, with an infinite number of a priori free parameters, which makes the theory intractable in full generality\footnote{In principle, the string effective action must also contain an infinite and well organized series of higher curvature corrections to the leading Einstein gravity \cite{Alvarez-Gaume:2015rwa}. However, any truncation of such action for practical calculations will bring the problems of the effective field theory back.}. In this situation, a sensible thing to do is to view the theory described by the action (\ref{oriact1}) as the effective theory applicable in the low energy/curvature regime only (see below for a more quantitative description). Adopting this approach to quadratic gravity (\ref{oriact1}) in this work we aim to infer constraints on $\beta$, and $\gamma$ parameters from the existing gravitational wave observations \cite{LIGOScientific:2019fpa}.

The study of quadratic gravity has a long history [for a recent review see \cite{Salvio:2018crh}, and references therein]. In fact, it has been known for a long time \cite{Stelle:1976gc} that quadratic gravity by itself while being a renormalisable theory contains, however, a fatal deficiency - propagating negative energy states, which compromises it's stability\footnote{See, e.g., Refs \cite{Hawking:2001yt, Bender:2007wu, Narain:2011gs, Salvio:2015gsi, Holdom:2015kbf, Raidal:2016wop, Donoghue:2018izj, Anselmi:2018bra, Salvio:2019ewf} for some alternative interpretations, however.}. More recently, gravitational waves generated by binary inspirals have also been studied. In \cite{Yagi:2011xp} dynamical quadratic gravity with scalar field dependent parameters has been studied with the main focus on the additional massless scalar waves. In \cite{Cao:2013osa} gravitational waveforms were studied in pure scalar quadratic gravity (with $\gamma=0$) with an additional assumption that the black hole binary is surrounded by the shell of a scalar field. In Ref. \cite{Calmet:2017rxl}, which is the closest to our present study, a constraint on the spin-2 wave mass (equivalently on $\gamma$) were obtained under the assumption that the massive spin-2 waves can reach the detector. In \cite{Caprini:2018oqe, Holscher:2019swu} the limits on the conformal versions of quadratic gravity has been discussed. Finally, in \cite{Holdom:2016nek} post-merger formation of the horizonless 2-2 hole has been considered with the characteristic prediction of gravitational echoes during the ringdown phase. None of these scenarios are feasible within the consistent effective field theory treatment of quadratic gravity adopted in the current work.

The rest of the paper is organised as follows. In Sec. \ref{sec2} we rewrite the action (\ref{oriact1}) in an equivalent and more convenient form separating out the massive spin-0, and the massive spin-2 fields, and will present their solutions in the linearised approximation. Given these solutions we compute the leading order corrections to the waveforms generated by a binary inspiral in Sec. \ref{sec3}. Constraints on quadratic gravity based on these calculations are discussed in Sec. \ref{sec4}. The final Sec. \ref{sec5} is reserved for our conclusions.

\subsection{Summary of Methodology}
It may be instructive to summarise our method and philosophy in slightly less technical language than as follows in the rest of this paper. First, we reinterpret the quadratic gravity degrees of freedom as massive scalar and tensor fields, within the realm of validity of an effective field theory for gravity. Since we expect these extra fields to be small compared to usual GR, we can find field equations and solutions for these fields to lowest order.

The post-Newtonian (PN) expansion scheme provides a method for perturbative calculations of the effects of gravity on bodies moving slow compared to the speed of light. This is certainly true for black hole binaries in their inspiral phase, and provides a useful framework to study modifications of gravity. As a perturbative scheme, the largest corrections to usual GR would occur at lowest PN order. Nominally, if one was to modify gravity, to place constraints on the modifications from the actual GW observations one would need to run a Bayesian analysis, varying all parameters to determine the maximum possible deviations from GR. However, this is computationally expensive, and technically difficult. In place of this sort of analysis, one can alternatively utilise the results of \cite{LIGOScientific:2019fpa}, in which they run a Bayesian analysis, varying all GR parameters to determine the maximum possible deviation from GR of any particular order of the waveform. Since this conservative bound involves varying \textit{all} parameters in the scenario, one would not expect that a modification of gravity could be "absorbed", for example, into a re-scaling of the black hole masses, or any other parameter. This provides us with an order-of-magnitude estimate of any possible modification of GR, which is convenient for our analysis, which is already to lowest-order, and can only aspire to find an order-of-magnitude constraint. One also sees here that the largest observational constraints on possible modifications exist at the lowest PN order, further justifying the decision to calculate corrections to lowest PN order, since this will provide the most robust constraints on any modification.

With the above in mind, we must then calculate the GW signal to lowest order. This is done by first deriving the motion of the two black holes, and in particular their phase as a function of time. In usual GR, the binary loses energy due to GW emissions. In our modification, additional scalar and tensor radiation affect the binary motion as well, which in turn straightforwardly modifies the GW waveform detected at Earth. The calculation of this waveform is conducted in the well-known stationary phase approximation. As explained above, conservative constraints on the possible deviation of this waveform from GR already exist, and these constraints can then be used to directly constrain our modification of gravity.

\section{Massive Scalar and Spin-2 Gravitons}\label{sec2}
In quadratic gravity (\ref{oriact1}) besides the massless spin-2 graviton, the theory contains additional propagating degrees of freedom: a massive scalar, $\phi$, and a massive spin-2 graviton, $\pi_{\mu\nu}$. It is convenient to explicitly separate these degrees of freedom by rewriting the quadratic action (\ref{oriact1}) in the form of the Einstein-Hilbert action supplemented by terms describing the extra massive degrees of freedom. This can be achieved by explicitly introducing $\phi$, and $\pi_{\mu\nu}$ fields in the action through the set of Lagrange multipliers which enforce the relations: $\phi=\frac{1}{2}R$, and $\pi_{\mu\nu}=R_{\mu\nu}-\frac{1}{4}g_{\mu\nu}R$. After field redefinitions and keeping terms quadratic in $\phi$, and $\pi_{\mu\nu}$ only, we arrive at the following action \cite{Tomboulis:1996cy}:
\begin{equation} \label{oriact2}
    S = \int d^4x \sqrt{-\tilde g} \left[\frac{\tilde{R}}{2\kappa} - \frac{1}{2} \left( \partial_{\mu} \phi \partial^{\mu} \phi + m^2_{\phi} \phi^2 \right) - \frac{1}{2} \left( \partial_{\mu} \pi^{\alpha \beta} \partial^{\mu} \pi_{\alpha \beta} + m^2_{\pi} \pi^{\alpha \beta} \pi_{\alpha \beta} \right) \right],
\end{equation}
where $\tilde R$ is the Ricci scalar constructed out of the metric
\begin{equation}\label{metric}
    g_{\mu \nu} = \tilde{g}_{\mu \nu} + \sqrt{\frac{2 \kappa}{3}} \eta_{\mu \nu} \phi + \sqrt{4 \kappa} \pi_{\mu \nu},
\end{equation}
to leading order in $\phi$ and $\pi_{\mu\nu}$, and the mass terms are defined as:
\begin{equation}\label{mass}
    m^2_{\phi} = \frac{M_P^2}{3(4\beta+\gamma)}, ~~m^2_{\pi} = \frac{M_P^2}{2 \gamma}.
\end{equation}

We require $\gamma \geq 0$, and $4\beta \geq -\gamma$ in order to avoid tachyonic instabilities. Notice, however, that the massive spin-2 field in (\ref{oriact2}) has the wrong sign for its kinetic term. To avoid the associated ghost instabilities, and to maintain theoretical consistency in our calculations we need to clarify the domain of validity of the effective quadratic gravity. This can be done by assuming that the quadratic action is a truncated asymptotic series expansion in powers of curvature tensor, where the error introduced by the truncation at a given power is smaller than the next terms in the expansion. Suppose we are interested in a system with a typical size $r$ (e.g., the size of a binary inspiral). Then $|R|\sim 1/r^2$ and the requirement that the Einstein-Hilbert term in (\ref{oriact1}) dominates over the quadratic terms implies  $M_P^2r^2/\beta, M_P^2r^2/\gamma > 1$, which can be translated into the constraints: $m_{\phi,\pi}r>1$.

There are few important ramifications of the above restrictions coming from the validity of quadratic truncation of the effective gravity action. First, it is clear from the action (\ref{oriact2}) that vacuum solutions of the Einstein gravity (e.g., the Schwarzschild black hole solution) is also a solution in quadratic gravity with trivial background massive fields $\phi = \pi_{\mu\nu} = 0$. The stability of these solutions within the domain of applicability of quadratic gravity are then guaranteed, and all the exotic non-perturbative solutions, such as the 2-2 hole solution, should be dismissed \cite{Simon:1990jn}. Second, theoretically, reliable calculations can be performed only in the regime with not too big curvature tensors, e.g. in the inspiral phase (this study), or in the ringdown phase. The merger phase is in principle intractable within the effective theory approach. Finally, the constraints $m_{\phi, \pi}r>1$, imply that the flux of massive waves cannot be produced during the inspiral phase. This comes from the requirement that the frequency of the waves arriving at the detector, given approximately by an inspiral angular velocity $\omega\simeq \Omega \approx v/r$, must exceed the mass, $\omega>m_{\phi, \pi}$ and this requirement is in contradiction with the constraints $m_{\phi, \pi}r>1$, since $v<1$ during the inspiral phase. Therefore we do not expect massive waves to be produced during the inspiral phase (see also \cite{Holscher:2018jhm}). We would like to stress that these conclusions are conservative and are purely based on the validity of quadratic gravity as an effective field theory. We do not exclude other treatments of quadratic gravity (perhaps supplemented by extra fields) where our constraints are relaxed.

The binary system during the inspiral phase is simply modelled as two point particles with masses $m_a$, and 4-velocities $v_a^{\mu}$, $a=1,2$. The action (in redefined fields) reads:
\begin{equation}\label{matter}
    S_{B}=\sum_{a=1}^{2}m_a\int dt\sqrt{-g_{\mu\nu}v_a^{\mu}v_a^{\nu}}.
\end{equation}

Next we consider linear perturbations, $\tilde h_{\mu\nu}=\tilde g_{\mu\nu}-\eta_{\mu\nu}$, $\phi$, and $\pi_{\mu\nu}$ and combining the actions (\ref{oriact2}), and (\ref{matter}), we obtain the following Euler-Lagrange equations:
\begin{align}
    \Box \Bar{\tilde{h}}_{\mu \nu} = - 8\pi G \sum^2_{a=1} m_a v_{\mu a} v_{\nu a} \delta^3(\vec{x} - \vec{y}_a(t)), \label{eom1} \\
    \Box \phi - m^2_{\phi} \phi = -  \sqrt{\frac{4\pi G}{3}} \sum^2_{a=1} m_a  \> \delta^3(\vec{x} - \vec{y}_a(t)), \label{eom2}  \\
    \Box \pi_{\mu \nu} - m^2_{\pi} \pi_{\mu \nu}= \sqrt{8\pi G} \sum^2_{a=1} m_a \left(v_{\mu a} v_{\nu a} + \frac{1}{4} \eta_{\mu \nu} \right) \delta^3(\vec{x} - \vec{y}_a(t)), \label{eom3}
\end{align}
where $\Bar{\tilde {h}}_{\mu\nu}=\tilde {h}_{\mu\nu}-\frac{1}{2}\eta_{\mu\nu}\tilde {h}^{\alpha}_{\alpha}$, and $\partial^{\mu}\Bar{\tilde {h}}_{\mu\nu}=\partial^{\mu}\pi_{\mu\nu}=\pi^{\mu}_{\mu}=0$. Eq. (\ref{eom1}) describes the usual massless gravity waves produced by the binary inspiral, while (\ref{eom2}), and (\ref{eom3}) describe  massive scalar and spin-2 waves, respectively. Since the massive waves contain scalar polarizations, they result in monopole and dipole radiation in addition to the standard quadrupole radiation already at the lowest order of the post-Newtonian (PN) expansion. Therefore, to constrain two parameters of quadratic gravity it is sufficient to compute corrections to the waveform at -2PN; however, we should note that observational constraints from gravitational wave experiments are currently only available at -1PN order. Higher-PN orders can also be systematically computed following the known formalism (see the review Ref. \cite{Blanchet:2013haa}, and references therein) in this framework. The solution for $\Bar{\tilde {h}}_{\mu\nu}$ is well known. The solutions of the massive field (\ref{eom2}), and (\ref{eom3}) can also be straightforwardly obtained using massive, retarded Green's functions:
\begin{align}
    \phi(x) = \sqrt{\frac{G}{12 \pi}}\sum^2_{a=1}  m_a  \frac{e^{-m_{\phi}  \left|\vec{x} - \vec{y}_a(t_r) \right|}}{\left|\vec{x} - \vec{y}_a(t_r) \right|}, \label{sol1} \\
    \pi_{\mu \nu}(x) = - \sqrt{\frac{G}{2 \pi}}\sum^2_{a=1}  m_a \left(v_{\mu a} v_{\nu a} + \frac{1}{4} \eta_{\mu \nu} \right) \frac{e^{-m_{\pi}  \left|\vec{x} - \vec{y}_a(t_r) \right|}}{\left|\vec{x} - \vec{y}_a(t_r) \right|}, \label{sol2}
\end{align}
where $t_r = t - \left|\vec{x} - \vec{y}_a(t_r) \right|$ is the retarded time. As it was expected, these solutions demonstrate that the massive fields mediate additional short-range Yukawa interactions in the binary system. Note that, while the interactions mediated by the massive scalar are attractive, the massive spin-2 generates a repulsive force. Of course, this is due to the fact that the $\pi_{\mu\nu}$ field describes negative energy states. Using these solutions, we are ready now to compute leading order corrections to the inspiral waveforms in quadratic gravity.

\section{Computing Leading Order Corrections to the Inspiral Waveforms}\label{sec3}
Our aim is to compute the leading order corrections to the phase of gravitational waves generated in the inspiral regime of a compact binary within quadratic gravity. To this end, we start with the covariant conservation equation,
\begin{equation}\label{conserv}
    \tilde \nabla_{\mu}\tilde T^{\mu\nu} = 0,
\end{equation}
where $\tilde \nabla_{\mu}$ is the covariant derivative constructed out of the metric $\tilde g_{\mu\nu}$ (\ref{metric}), and the energy momentum tensor of the binary is defined as:  
\begin{equation}\label{emt}
    \tilde T^{\mu\nu} = \frac{-2}{\sqrt{-\tilde g}}\frac{\delta S_B}{\delta \tilde g_{\mu\nu}}=\sum_{a=1}^2 m_a\frac{v_a^{\mu}v_a^{\nu}}{\sqrt{-\tilde g_{\mu\nu}v_a^{\mu}v_a^{\nu}}}\delta^3\left(\vec x-\vec y_a(t)\right).
\end{equation}

Picking up $\nu=i$ spatial components in (\ref{conserv}), and integrating over the spatial volume surrounding only one of the two compact objects, we obtain the modified Newton's force law describing the dynamics of the object:
\begin{equation}\label{eom0}
    \frac{d P^i_a}{dt}=F^i_a+\sqrt{\frac{16\pi G}{3}}\partial^i\phi+\sqrt{32\pi G}\partial^i\pi_{\mu\nu}v_a^{\mu}v_a^{\nu},
\end{equation}
where $P^i_a$ is the linear momentum of the $a$-th object, and $F^i_a$ is the force acting on it due to the massless graviton exchange \cite{Blanchet:2000ub}:
\begin{equation}
    P^i_{a} =  \frac{[\tilde g_{\mu i}]_a v^{\mu}_a}{\sqrt{- [\tilde g_{\rho \sigma}]_a v^{\rho}_a v^{\sigma}_a}}, ~~ F^i_{a} = \frac{1}{2} \frac{\partial_i [\tilde g_{\mu \nu}]_a v^{\mu}_a v^{\nu}_a}{\sqrt{-[\tilde g_{\rho \sigma}]_a v^{\rho}_a v^{\sigma}_a}}.
\end{equation}

From (\ref{eom0}), and using our solutions (\ref{sol1}, \ref{sol2}) we calculate the accelerations, $\vec a_{1,2}=\frac{d\vec v_{1,2}}{dt}$, as:
\begin{equation}\label{accNew}
    \vec a_1 = -\frac{G m_2}{r^2_{12}} \hat{n}_{12} \left(1 + \frac{2}{3} e^{- m_{\phi} r_{12}} \left(m_{\phi} r_{12} + 1 \right) -3e^{- m_{\pi} r_{12}} \left(m_{\pi} r_{12} + 1 \right) \right),
\end{equation}
where $r_{12} = \left|\vec{y}_1 - \vec{y}_2 \right|$, and $\hat{n}_{12} = (\vec{y}_1 - \vec{y}_2)/r_{12}$ ($\vec a_2=\vec a_1 [1\leftrightarrow2]$). As we are interested in relative motion, we introduce convenient parameters: the total mass $M=m_1+m_2$, the reduced mass $\mu=\frac{m_1m_2}{m_1+m_2}$, and the symmetric ratio $\nu=\frac{\mu}{M}$.
The relative acceleration, $\vec a=\vec a_1-\vec a_2$ then reads:
\begin{equation}\label{relaccNew}
    \vec a = -\frac{G M}{r^2} \hat{n} \left(1 + \frac{2}{3} e^{- m_{\phi} r} \left(m_{\phi} r + 1 \right) -3e^{- m_{\pi} r} \left(m_{\pi} r + 1 \right) \right),
\end{equation}
where $\hat{n} = \hat{n}_{12} = -\hat{n}_{21}$, dropping the subscript.

In the quasi-circular orbit approximation this defines the angular frequency of the binary system:
\begin{align}\label{angfreq}
    \Omega^2 &= \frac{G M}{r^3} \left(1 + \frac{2}{3} e^{- m_{\phi} r} \left(m_{\phi} r + 1 \right) -3e^{- m_{\pi} r} \left(m_{\pi} r + 1 \right) \right) \nonumber \\
    &\equiv \frac{G M}{r^3} A.
\end{align}

The total energy of the virialised binary is given by:
\begin{equation}\label{conenevir}
    E = -\mu \frac{G M}{r} \left(\frac{1}{2} + \frac{2}{3} e^{- m_{\phi} r} -3e^{- m_{\pi} r} \right).
\end{equation}

We see that the massive scalar field decreases the energy of the binary, while the massive spin-2 carries out negative energy, and increases the energy of the binary system. The rate of the energy change is convenient to calculate by introducing a dimensionless frequency-related parameter: 
\begin{equation}\label{postnewpar}
    x \equiv \left(G M \Omega \right)^\frac{2}{3}.
\end{equation}

The relative distance in terms of $x$ reads:
\begin{equation}\label{rgx}
    r = \frac{G M}{x} A^{\frac{1}{3}}.
\end{equation}

Using (\ref{rgx}) we replace $r$ by $x$ in (\ref{conenevir}) keeping leading order terms, and taking the derivative, we obtain:
\begin{align}\label{dconenex}
    \frac{d E}{d x}& = -\mu  \biggl[\frac{1}{2} + \frac{1}{9} e^{- m_{\phi} \frac{G M}{x}} \left(5 + 5 m_{\phi} \frac{G M}{x} - \left(m_{\phi} \frac{G M}{x} \right)^2 \right) \nonumber \\
    &- \frac{1}{2} e^{- m_{\pi} \frac{G M}{x}} \left(5 + 5 m_{\pi} \frac{G M}{x} - \left(m_{\pi} \frac{G M}{x} \right)^2 \right) \biggr].
\end{align}

The change of the energy of binary must be balanced by the flux of emitted gravitational waves, hence:
\begin{equation}\label{balequ1}
    \frac{d E}{d t} = - \mathcal{F}.
\end{equation}

We have argued previously, that the massive scalar, and the massive spin-2 waves cannot contribute to the radiated flux, therefore $\mathcal{F}$ in (\ref{balequ1}) is given by the standard GR expression:
\begin{equation}\label{flux}
    \mathcal{F} = \frac{32}{5 G} \nu^2 x^5. 
\end{equation}

We now have most of the tools we need to calculate the binary phase. First we introduce the dimensionless time variable $\Theta$:
\begin{equation}\label{theta}
    \Theta \equiv \frac{\nu}{5 G M} \left(t_c - t \right),
\end{equation}
where $t_c$ is the instant of coalescence. The orbital phase $\varphi$ of the binary system is defined as $d \varphi / d t = \Omega$. This in terms of $\Theta$ reads:
\begin{equation}\label{phathe}
    \frac{d \varphi}{d \Theta} = - \frac{5}{\nu} x^{3/2}.
\end{equation}

Using (\ref{theta}), and (\ref{phathe}) we can rewrite the energy balance equation as
\begin{equation}\label{balequ2}
    \frac{d E}{d x} \frac{d x}{d \varphi} \frac{d \varphi}{d \Theta} \frac{d \Theta}{d t} =\frac{d E}{d x} \frac{d x}{d \varphi} \frac{x^{3/2}}{G M} = - \mathcal{F}.
\end{equation}

Next, using (\ref{dconenex}), and (\ref{flux}) the we obtain the differential equation for the phase,
\begin{align}\label{phadiff}
    \frac{d \varphi}{d x} = \frac{5 x^{-7/2}}{32 \nu} \biggl[\frac{1}{2} &+ \frac{1}{9} e^{- m_{\phi} \frac{G M}{x}} \left(5 + 5 m_{\phi} \frac{G M}{x } - \left(m_{\phi} \frac{G M}{x } \right)^2 \right) \nonumber \\
    &- \frac{1}{2} e^{- m_{\pi} \frac{G M}{x}} \left(5 + 5 m_{\pi} \frac{G M}{x} - \left(m_{\pi} \frac{G M}{x} \right)^2 \right) \biggr],
\end{align}
which can be easily integrated: 
\begin{align}\label{phase}
    \varphi = -\frac{x^{-5/2}}{32 \nu} \biggl[1 &+ e^{- m_{\phi} \frac{G M}{x}} \left(\frac{5}{6} - \frac{5}{9} m_{\phi} \frac{G M}{x} \right) 
    - e^{- m_{\pi} \frac{G M}{x}} \left(\frac{15}{4} - \frac{5}{2} m_{\pi} \frac{G M}{x} \right)  \biggr].
\end{align}

This is the desired result which describes the leading corrections to the binary inspiral phase. The exponentially suppressed terms in (\ref{phase}) are due to additional massive scalar, and massive spin-2 degrees of freedom in quadratic gravity. Among those, corrections $\propto x^{-7/2}$ are due to the Yukawa suppressed dipole-like contributions, which is totally absent in GR, and can be classified as -1PN corrections. The corrections $\propto x^{-5/2}$ come from Yukawa quadrupole radiation and are at 0PN Newtonian order in the post-Newtonian expansion.

\section{Constraints on Quadratic Gravity}\label{sec4}
Previous constraints on quadratic gravity parameters are based on the modification of the Newtonian inverse square law introduced by the exchange of the massive scalar, and the massive spin-2 mediators. This can be probed in various experiments, including tabletop torsion balance \cite{Hoyle:2004cw}, and laser ranging satellite experiments \cite{Lucchesi:2018cdl, Merkowitz:2010kka}. While significant constraints on the strength of Yukawa modification of the Newtonian potential are obtained in those experiments, which invalidates quadratic gravity, one must keep in mind that that those constraints are obtained only for specific ranges of interactions. For example, the torsion balance experiments excludes quadratic gravity only for $10^{-5}\text{eV}\lesssim m_{\pi,\phi}\lesssim 10^{-3}\text{eV}$ \cite{Hoyle:2004cw} (in terms of the parameters, using (\ref{mass}), $2.9\cdot 10^{60} \leq \gamma \, \lesssim 2.9\cdot 10^{64}$, and $4.8\cdot 10^{59} -\frac{\gamma}{4} \leq \beta \, \lesssim 4.8\cdot 10^{63} - \frac{\gamma}{4}$), while satellite experiments are sensitive only to the range comparable to the orbital size they monitor: $m_{\pi,\phi}\simeq 3\cdot 10^{-14}$ eV (earth ranging \cite{Lucchesi:2018cdl}), and  $m_{\pi,\phi}\simeq 5\cdot 10^{-16}$ eV (lunar ranging \cite{Merkowitz:2010kka}).

On the contrary, bounds on quadratic gravity that can be inferred from the modification of the gravitational waveform is \emph{a priori} valid for an arbitrary range and thus we expect genuine upper limits, which we will derive now. The gravitational wave strain signal with amplitude $A(t)$, and phase $\Phi(t)$ takes the following form:
\begin{equation}\label{gwampha}
    h(t) = 2 A(t)\cos{\Phi(t)}.
\end{equation}

Since the gravitational wave frequency is twice the orbital frequency of the binary system we have $\Phi=2\varphi$. Next, using (\ref{balequ1}) we find $x$ as a function of the dimensionless time parameter $\Theta$,
\begin{equation}\label{xtheta} 
    x = \frac{1}{4} \Theta^{-1/4} \left[1 + \frac{2}{9} e^{-4 m_{\phi} G M \Theta^{1/4}} \left(4 m_{\phi} G M \Theta^{1/4} \right) - e^{-4 m_{\pi} G M \Theta^{1/4}} \left(4 m_{\pi} G M \Theta^{1/4} \right) \right],
\end{equation}
and rewrite (\ref{phase}) as an explicit function of time:
\begin{align}\label{phase2}
    \varphi = -\frac{\Theta^{5/8}}{\nu} \biggl[1 &+ e^{-4 m_{\phi} G M \Theta^{1/4}} \left(\frac{5}{6} - \frac{40}{9} m_{\phi} G M \Theta^{1/4} \right) \nonumber \\
    &- e^{-4 m_{\pi} G M \Theta^{1/4}} \left(\frac{15}{4} - 20 m_{\pi} G M \Theta^{1/4} \right) \biggr].
\end{align}

The gravitational wave forms are analysed in the frequency space, however. Fourier transform of (\ref{gwampha}) in the stationary phase approximation \cite{Damour:2000zb} reads:
\begin{align}
    h(f) &= \frac{\sqrt{2\pi}A(t_f)}{\sqrt{\ddot{\Phi}(t_f)}}e^{i\psi(f)},\\
\end{align}
where the parameter $t_f$ is given by the time when the gravitational wave frequency $d \Phi(t)/dt$ is equal to the Fourier frequency $f$. The phase can be explicitly evaluated and written as expansions in PN orders: 
\begin{equation}
    \psi(f) = 2 \pi f t_c - \phi_c - \pi/4 + \frac{3}{128 \nu} \sum^0_{i=-4} \varphi_i (G M \pi f)^{(i-5)/3},
\end{equation}
where $t_c$, and $\phi_c$ are the time, and phase at coalescence respectively. In particular, we obtain that the highest order corrections to the standard GR phase to be of negative second, and negative first order PN corrections:
\begin{align}\label{freqcoeff}
    -2\text{PN} \quad \varphi_{-4} &= \frac{106336}{81} (m_{\phi} G M)^2 e^{-m_{\phi} \delta} - \frac{53168}{9} (m_{\pi} G M)^2 e^{-m_{\pi} \delta}, \nonumber \\
    -1\text{PN} \quad \varphi_{-2} &= \frac{451928}{81} m_{\phi} G M e^{-m_{\phi} \delta} - \frac{225964}{9} m_{\pi} G M e^{-m_{\pi} \delta},
\end{align}
where $\delta=4 G M \left(8 \pi G M f \right)^{-2/3}$. The advantage of using gravitational waves to constrain quadratic gravity steams from the fact that the leading correction arises at lower than Newtonian order (see, (\ref{phase})). The absolute deviation of the -1PN phase has been constrained from observations of gravitational waves \cite{LIGOScientific:2019fpa} to be $|\delta \phi_{-1PN}|< 10^{-2}$, while (to our knowledge) there does not exist at the moment -2PN constraints \cite{LIGOScientific:2019fpa}. From our calculations of -1PN correction (\ref{freqcoeff}), we then obtain:
\begin{equation}\label{simequ2}
    \left|\frac{451928}{81} m_{\phi} G M e^{-m_{\phi} \delta} - \frac{225964}{9} m_{\pi} G M e^{-m_{\pi} \delta} \right| \lesssim 10^{-2}.
\end{equation}

We could also write explicitly higher order corrections to the GR phase. However, the lower PN bounds provide tighter constraints on quadratic gravity than higher PN bounds and we do not display higher PN corrections here. Assuming no accidental cancellation between the two terms on the left hand side of (\ref{simequ2}) and taking the typical values $f=75$Hz, and $M=30 M_\odot$, the inequality is reduced to:
\begin{equation}
    m_{\phi,\pi} G M e^{- m_{\phi, \pi} \delta}\lesssim 6.0\cdot 10^{-7},
\end{equation}
which leads to the bound:
\begin{equation}\label{limits1} 
    m_{\phi,\pi} \gtrsim 7.0\cdot 10^{-12}~\text{eV}.
\end{equation}

At last, using (\ref{mass}), we rewrite (\ref{limits1}) as limits on the original dimensionless parameters of quadratic gravity:
\begin{align}
    0 \leq & \enspace \> \gamma \enspace \, \lesssim 5.9\cdot 10^{76}, \label{limits2} \\
    -\frac{\gamma}{4} \leq & \enspace \> \beta \enspace \, \lesssim 9.8\cdot 10^{75} - \frac{\gamma}{4}. \label{limits3}
\end{align}

The bound on $\beta$ (properly translated) is 4 orders of magnitude stronger than the one anticipated in \cite{Cao:2013osa}.

As a final remark we note that ideally we would use -2PN correction to the GW waveform to make our constraints, since throughout our analysis we have been taking lowest-order terms only. This means the exact values of the -1PN coefficients given above should actually be modified; for example, the multiplication of a -2PN term with a 1PN term would result in a -1PN term which needs to have been kept for full accuracy. Our calculation is only tractable, however, by remaining at lowest order. One would not expect the order of magnitude of the coefficients to vary greatly in the full solution, so our order-of-magnitude estimation above is safe. Once the -2PN constraints are known, we could instead use those to more accurately place bounds on quadratic gravity.

\section{Conclusions}
\label{sec5}
In this paper, we have studied gravitational wave signals from the inspiral phase of a compact binary system within the framework of quadratic gravity. Quadratic gravity has been considered as a truncated approximation of a theory represented by an action written in terms of the powers of curvature tensors. Such an action necessarily emerges as a result of renormalisation of the Einstein-Hilbert action within the semi-classical gravity at any fixed order of the quantum loop-expansion. Since semi-classical gravity is not renormalisable there are in principle an infinite number of free parameters corresponding to an infinite series of higher curvature terms. Therefore, in the truncated quadratic theory which contains only two parameters, we must make sure that we are working in the regime, where higher than quadratic curvature terms are sub-dominant. This restricts our consideration to inspiral phase of the evolution of the binary system. We have also argued that non-perturbative solutions significantly deviating from the GR solutions (e.g., the 2-2 hole solution) are not valid within our approximation. Furthermore, constraints on the theoretical validity of quadratic gravity, imply that no flux of additional massive radiation is produced during the inspiral phase.

Under these constraints, we have computed the leading corrections to the phase of the GR waveform due to the quadratic terms (\ref{phase}). These corrections comprise of formally -1PN corrections due to the dipole radiation of massive scalar, and scalar polarization of the massive spin-2 fields. Confronting these corrections with the existing data \cite{LIGOScientific:2019fpa}, we have extracted bounds (\ref{limits2}, \ref{limits3}) on both $\beta$, and $\gamma$ parameters of quadratic gravity. To the best of our knowledge these are the stringent limits available in the literature. 

Higher PN order calculations can in principle be performed. However, no improvement of the obtained bounds (\ref{limits1}, \ref{limits2}, \ref{limits3}) within quadratic gravity is expected\footnote{One should note, however, that beyond the classical calculations, effects of the virtual $\phi$, and $\pi$ states could contribute to higher PN terms without Yukawa suppression. It would be interesting to explicitly compute these corrections.}. On the other hand, one can envisage the program which probes higher than quadratic curvature terms by more accurate PN calculations of the binary inspiral. Ultimately, the best probe of higher curvature gravity must come from the signal produced during the merger phase. However, besides complications related with numerical calculations, theoretical control of the validity of finite truncation in merger phase becomes a significant challenge.

\subparagraph{Acknowledgement} We would like to thank Cyril Lagger for useful discussions.


\end{document}